\documentclass[onecolumn,pdftex]{revtex4}

\usepackage[pdftex]{graphicx}
\usepackage{amsmath}

\begin{document}

\title{Mass composition of cosmic rays above 0.1 EeV by the Yakutsk array data}

\author{S. Knurenko}
\author{I. Petrov$^{*}$}

\affiliation{Yu.G. Shafer Institute of Cosmophysical Research and Aeronomy SB RAS, 31 Lenin ave. 677980 Yakutsk, Russia}

\email{igor.petrov@ikfia.ysn.ru}

\begin{abstract}
The paper presents the results of the longitudinal development of extensive air showers (X$_{max}$) of ultra-high energies and mass composition of cosmic rays. The measurements of X$_{max}$ are based on data from observations of the Cherenkov radiation at the Yakutsk array for the period 1974-2014. The cascade curves of individual showers and the depth of maximum X$_{max}$ were reconstructed over the energy range 10$^{16}$-5.7$\cdot$10$^{19}$ eV. It is shown that the displacement rate of the parameter dX$_{max}$ / dE in the atmosphere is nonlinear and depends on the energy. Such a feature indicates a change in mass composition, which is confirmed by fluctuations of X$_{max}$ in this energy region. The composition of cosmic rays was determined by interpolation using the QGSJetII-04 model.
\end{abstract}


\maketitle


\section{Introduction}

Knowledge of the mass composition (MC) of cosmic rays (CR) in the energy range above 0.1 EeV is very important for astrophysics. It is assumed that a transition from galactic to extra-galactic CR \citep{Berezhko201231} is within the energy range of 0.1-1 EeV, also irregularities like "dip-bump" at energies above 5 EeV \citep{Berezinsky2006043005}, and cutoff of the spectrum of CR associated with the GZK effect at energies $\sim$50 EeV \citep{Greisen1966748,Zatsepin196678}. 

One of the crucial techniques in the study of the characteristics of CR by Extensive Air Shower (EAS) method is Vavilov-Cherenkov radiation \citep{Zrelov1968274, Jelley1958331} detection, which is produced by shower particles in the visible wavelength. The total flux of Cherenkov light allows determining the energy of the primary particle and the lateral distribution at sea level provides information on the nature of the development of a shower in the atmosphere \citep{Knurenko2001157}.

Longitudinal development of the shower in the atmosphere was reconstructed by the lateral distribution function (LDF) of Cherenkov light \citep{Knurenko2001157,Dyakonov1986} using the inverse problem method \citep{Tikhonov1977258}. To estimate the mass composition, the depth of maximum of air shower development X$_{max}$ was used, which is sensitive to the atomic mass of the primary particle.

At the Yakutsk array, in order to estimate air shower energy we used the energy balance method \citep{Knurenko2006473}. The method is based on the energy dissipated by EAS particles in the atmosphere, i.e. full flux of Cherenkov light. In individual showers, the energy was determined by the flux density of Cherenkov light at a distance R = 200 m from the shower axis - Q (200). According to calculations, the classification parameter Q (200) is proportional to the shower energy and have little shower-to-shower fluctuations. 

The average depth of the air shower maximum X$_{max}$ was used to derive CR MC within the hadronic interaction model QGSJetII-04 \citep{Ostapchenko201183} according to \citep{Horandel200641}. 

\begin{figure}[h]
	\center{\includegraphics[width=0.8\linewidth]{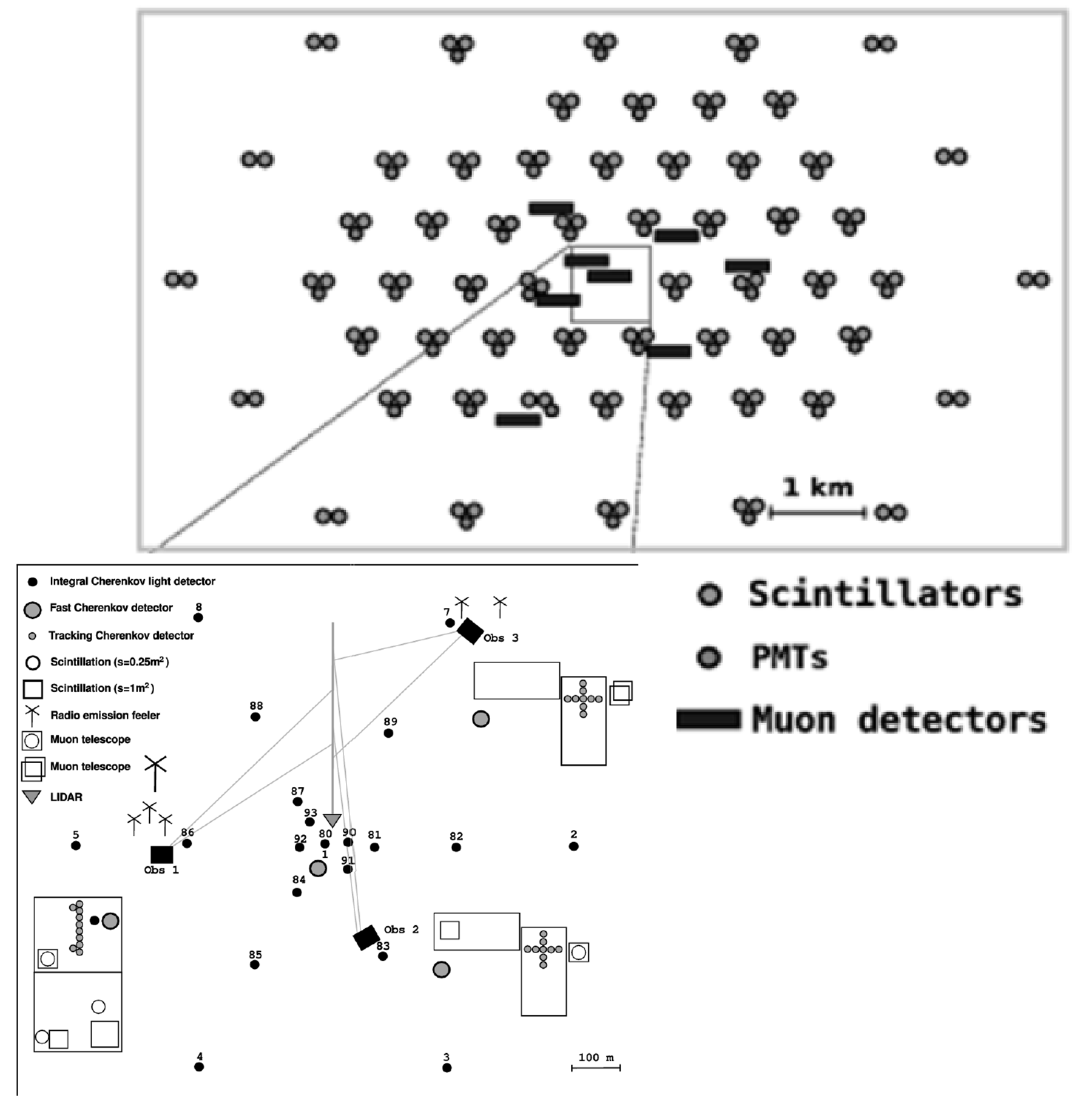}}
	\caption{Arrangement of the observation stations at the Yakutsk array}
	\label{fig_ykt_detector_arrangement}
\end{figure}

The Yakutsk array consists of stations with scintillation and Cherenkov detectors to register elementary particles of different nature: hadrons, electrons and positrons, muons, and Cherenkov photons \citep{Artamonov199412}. Additionally, there are surface electric field and air shower radio emission observations \citep{Knurenko2017230}. The array consists of 120 scintillation detectors located on an area of $\sim$13 km$^{2}$. Three muon stations and 72 channels for integrated and differential Cherenkov detectors are located at the center \citep{Knurenko199846,Garipov2001885,Ivanov201534,Egorov2018462} (Fig. \ref{fig_ykt_detector_arrangement}). The observation energy range of the array is from 10$^{15}$ eV to 10$^{20}$ eV, which allows it to study both galactic and extra-galactic CR. 

Cherenkov measurements are conducted on clear and moon-less nights. The observation season starts in September and ends in April. The total observation time per season is 450-600 hours, which takes 6-10$\%$ of the annual array operation time. Since optical observations are very sensitive to such atmospheric characteristics as transparency and the aerosol component of the atmosphere \citep{Dyakonov1999315,Zuev1970215}, the atmosphere is monitored daily at the array \citep{Dyakonov1991868,Knurenko20066522,Knurenko20149292}.

\section{Experimental data and analysis}
\subsection{Reconstruction of air shower parameters }
The depth of maximum X$_{max}$ is determined by the cascade curve of Cherenkov light, which is reconstructed from the experimental Cherenkov LDF (Fig. \ref{fig_ykt_ldf}) \citep{Knurenko2001157}. This algorithm is described in detail in \citep{Knurenko2001157,Dyakonov1986}. The algorithm is based on the Fredholm equation of the first kind (\ref{yakutsk_eq1}), which is solved by the adaptive method \citep{Kochnev198562}:

\begin{figure}[h]
	\center{\includegraphics[width=0.8\linewidth]{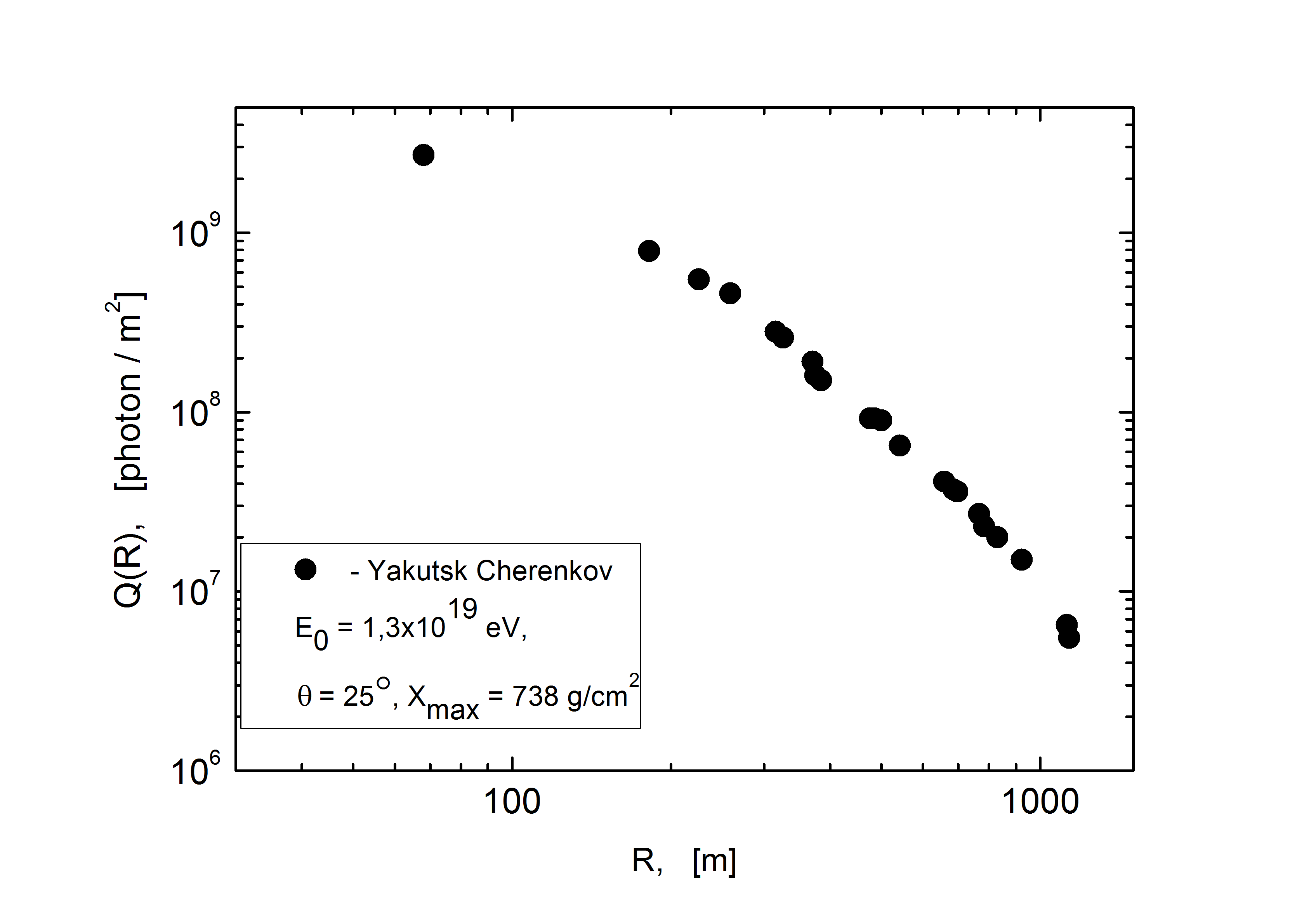}}
	\caption{LDF of the Cherenkov light}
	\label{fig_ykt_ldf}
\end{figure}

\begin{equation}\label{yakutsk_eq1}
Q_{exp} = \delta_{Q} + \int_{X_{1}}G(R,X/X_{2})\cdot N(E_0, X)\cdot K(\lambda, X)dX
\end{equation}

\begin{figure}[h]
	\center{\includegraphics[width=0.8\linewidth]{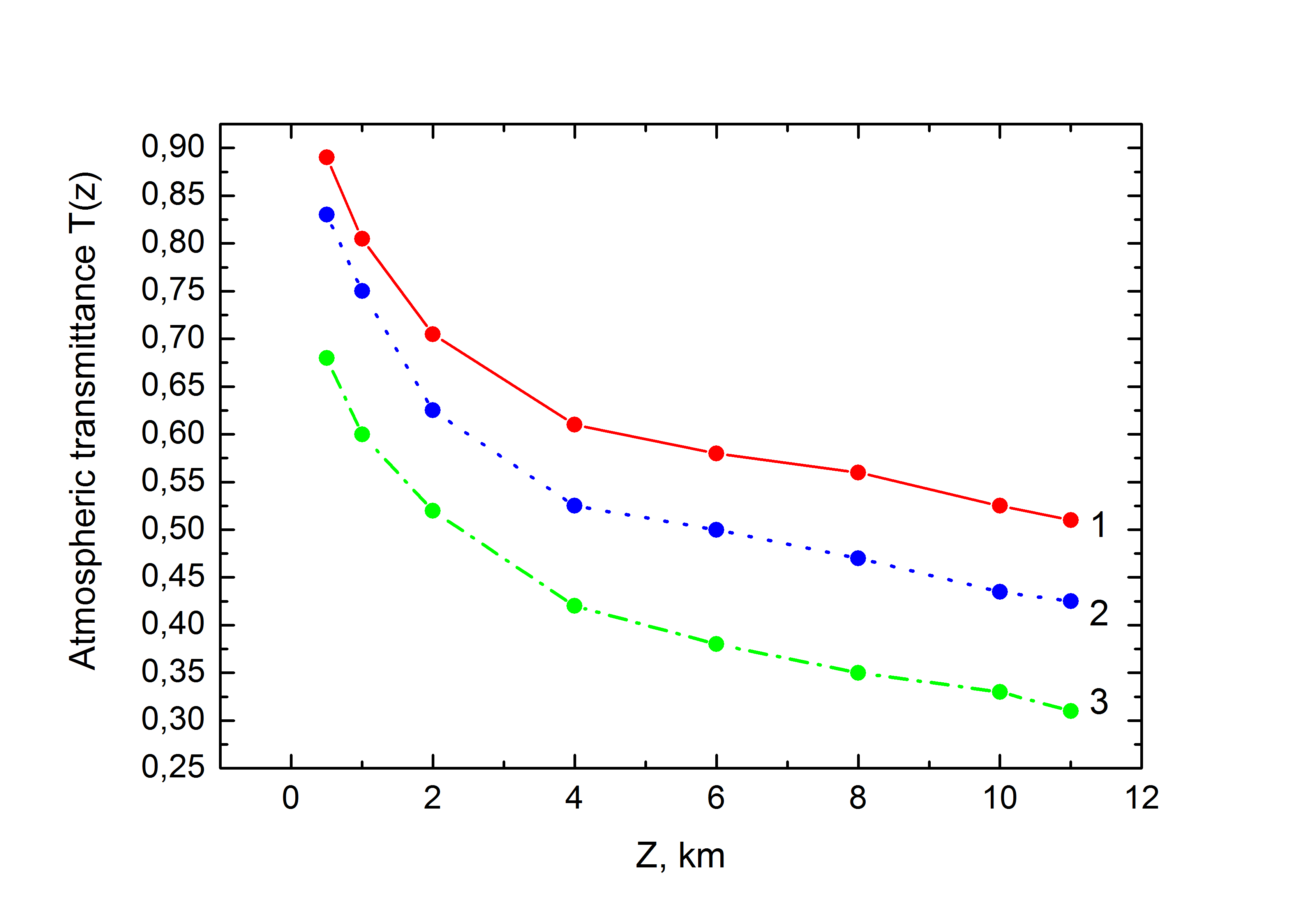}}
	\caption{Vertical profile of atmospheric transmittance coefficient for different conditions: 1- 5 points; 2- four points; 3- three points \citep{Dyakonov1999315}}
	\label{fig_ykt_transm}
\end{figure}

In equation (\ref{yakutsk_eq1}), $G(R,X/X_2)$ is the function that represents the lateral-angular distribution of electrons in the electron-photon cascade. In our case, the function $G(R,X/X_2)$ takes into account the nonequilibrium spectrum of electrons \citep{Belyaev1980305, Dyakonov1982151}, i.e. depends on the age of the shower and is close to the data from Hillas \citep{Hillas19828, Patterson19839} and Giller \citep{Giller200430}. More recent work \citep{Nerling200624,Smialkowski2018854} confirms the correctness of the use of lateral-angular distribution that depends on the age of the shower, for X$_{max}$ reconstruction by Cherenkov light of air showers. The use of different models of lateral-angular distribution from the "equilibrium" approximation and, including calculations of the electron spectra depending on shower age, yielded 3-5$ \% $ in the spread of the X$_{max}$ estimation. N (E$_{0}$, X) is a cascade curve, $\delta$Q is the "noise" level, depends on measurement uncertainty and data processing errors. The choice of lateral-angular distribution was made in \citep{Knurenko2003168d}. K ($\lambda $, X) is transmittance of the atmosphere that takes into account the transmission of Cherenkov light throughout the entire thickness of the atmosphere. The coefficient K ($\lambda $, X) was reconstructed by solving the inverse problem using the lateral distribution of Cherenkov light registered under different weather conditions by a photomultiplier with a maximum sensitivity at a wavelength of $ \lambda $ = 430 nm \citep{Dyakonov1999315}. Reconstructed dependence of Cherenkov light transmission coefficient for different weather conditions on the height is shown in Fig. \ref{fig_ykt_transm}. Approximation of these data is also given there.  

For the prompt estimation of atmospheric conditions during the registration of air showers in the optical wavelength at the Yakutsk array, a point system for estimation of the state of the atmosphere was developed. The estimation was tested by instrumental methods, including the rate of triggers for showers with energies of 10$^{15}$-10$^{16}$ eV \citep{Dyakonov1991868}. The point system for classification of the atmosphere was proposed in the papers \citep{Knurenko20066522, Guschin1988200} based on lidar measurements of the spectral transparency P$_{\lambda}$ and spectral atmospheric optical thickness $\delta_{\lambda} $. The correlation results for the three wavelengths are given in Table \ref{yakutsk_tab01}.  

\begin{table}
    \centering
	\caption{Different values for P$_{\lambda} $ and $\delta_{\lambda}$ in different scales of the atmosphere transparency.}
	\label{yakutsk_tab01}       
	\begin{tabular}{|c|c|c|c|c|c|}
		\hline 
		\rule[-1ex]{0pt}{2.5ex} Atmospheric transparency & P$_{\lambda=530} $ & $ \delta_{\lambda=530} $ & P$ _{\lambda=369} $ & $ \delta_{\lambda=369} $ & P$_{\lambda=430} $ \\ 
		\hline 
		\rule[-1ex]{0pt}{2.5ex} low (2 points) & 0.60 & 0.180 & 0.35 & 0.260 & 0.48 \\ 
		\hline 
		\rule[-1ex]{0pt}{2.5ex} decreased (3 points) & 0.70 & 0.080 & 0.46 & 0.120 & 0.56 \\ 
		\hline 
		\rule[-1ex]{0pt}{2.5ex} normal (3 - 4 points) & 0.75 & 0.060 & 0.50 & 0.080 & 0.60 \\ 
		\hline 
		\rule[-1ex]{0pt}{2.5ex} increased (4 points) & 0.80 & 0.040 & 0.52 & 0.060 & 0.68 \\ 
		\hline 
		\rule[-1ex]{0pt}{2.5ex} high (5 points) & 0.85 & 0.020 & 0.56 & 0.030 & 0.72 \\ 
		\hline 
	\end{tabular} 	
\end{table}

X$_{1}$ and X$_{2}$ - upper and lower limits of the atmosphere for air shower development. The limits were used for reconstruction of the air shower cascade curve.  

\begin{figure}[h]
	\center{\includegraphics[width=0.8\linewidth]{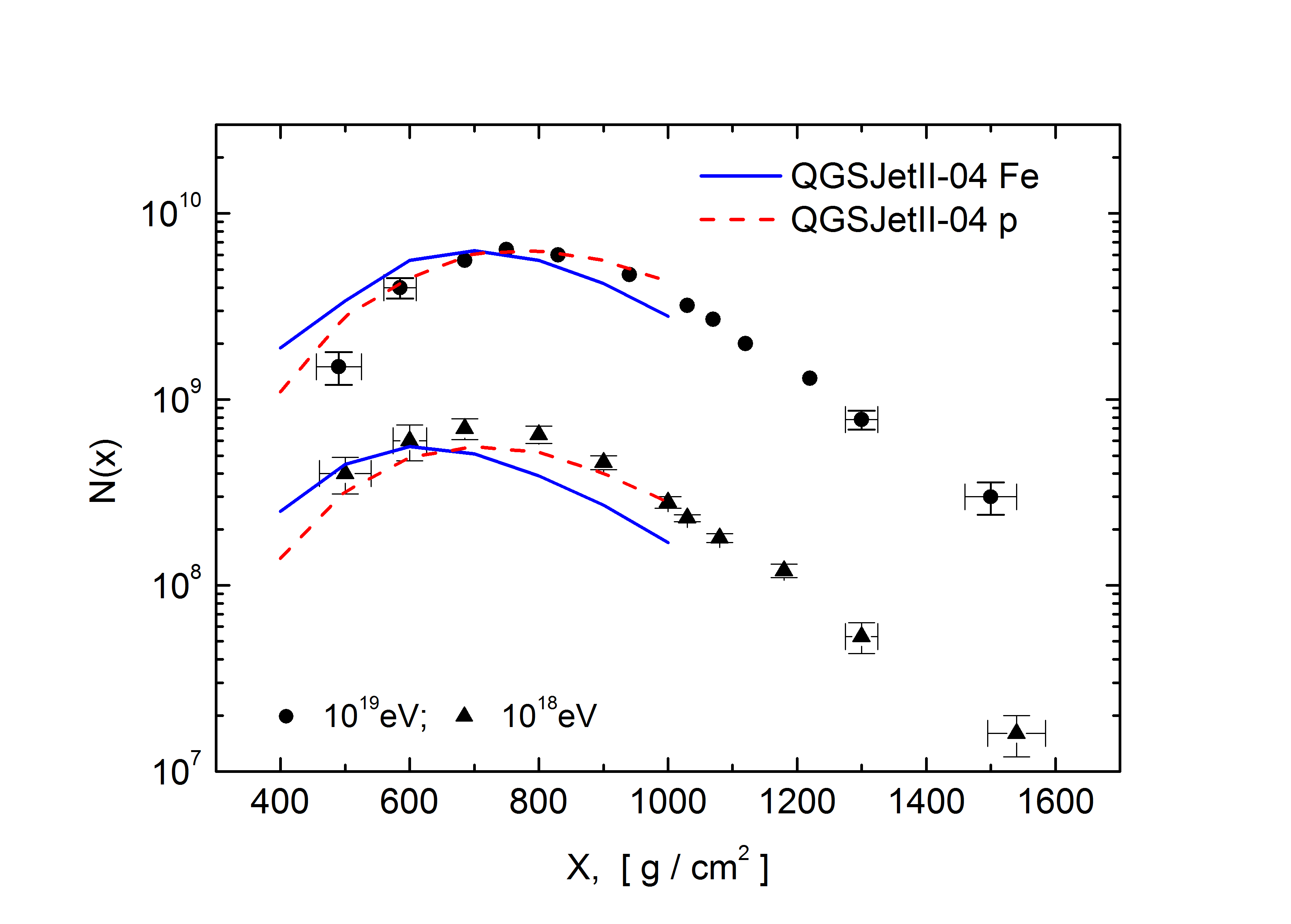}}
	\caption{Average experimental cascade curves and calculations by QGSJETII-04 \citep{Ostapchenko201183} for proton and iron}
	\label{fig_ykt_cascade_curve}
\end{figure}

As seen from equation (\ref{yakutsk_eq1}), the method takes into account the physics of air shower electron-photon component development and characteristics of the atmospheric conditions during registration of the Cherenkov radiation \citep{Knurenko20066522,Knurenko20149292} at the Yakutsk array. Using this technique, average cascade curves were reconstructed (Fig. \ref{fig_ykt_cascade_curve}) for energies of  $\sim$10$^{18}$ and $\sim$10$^{19}$ eV \citep{Tikhonov1977258} and compared with QGSJetII-04 simulations. As can be seen from Fig. \ref{fig_ykt_cascade_curve}, the experimental cascade curve is measured in a wide range of depths due to inclined showers. This circumstance was used to determine the absorption range of cascade particles beyond the maximum development of air showers \citep{Glushkov199357}. On the other hand, cascade curves can be used to test different models of the development of air shower in the atmosphere. 

The energy at the Yakutsk array was determined by energy balance method of all air shower particles \citep{Knurenko2006473}. This suggests that the total energy of the shower takes into account the ionization losses of each of the components of air shower: electrons, muons, hadrons and neutrinos (see formula (\ref{yakutsk_eq2})).

\begin{equation}\label{yakutsk_eq2}
E_0 = E_{ei} + E_{el} + E_\mu + E_{hi} + E_{\mu i} + E_\nu
\end{equation}

Their share of energy was determined empirically taking into account the experiment conducted at the Yakutsk array. For example, the ionization losses of the electronic component were determined by the total flux of Cherenkov light from air showers, according to formula (\ref{yakutsk_eq3}):

\begin{equation}\label{yakutsk_eq3}
E_{ei} = k(x,P_\lambda)\cdot\Phi
\end{equation}

where $\Phi$ is the total flux of Cherenkov light; k (x, P$_{\lambda}$) is the approximation coefficient (calculated value), takes into account the transparency of the real atmosphere, the nature of the longitudinal development of the shower (energy spectrum of secondary particles and its dependence on the age of the shower) and expressed through the depth of the maximum of air shower X$_{max}$, measured at the array \citep{Knurenko2006473,Ivanov201534}.

\begin{figure}[h]
	\center{\includegraphics[width=0.8\linewidth]{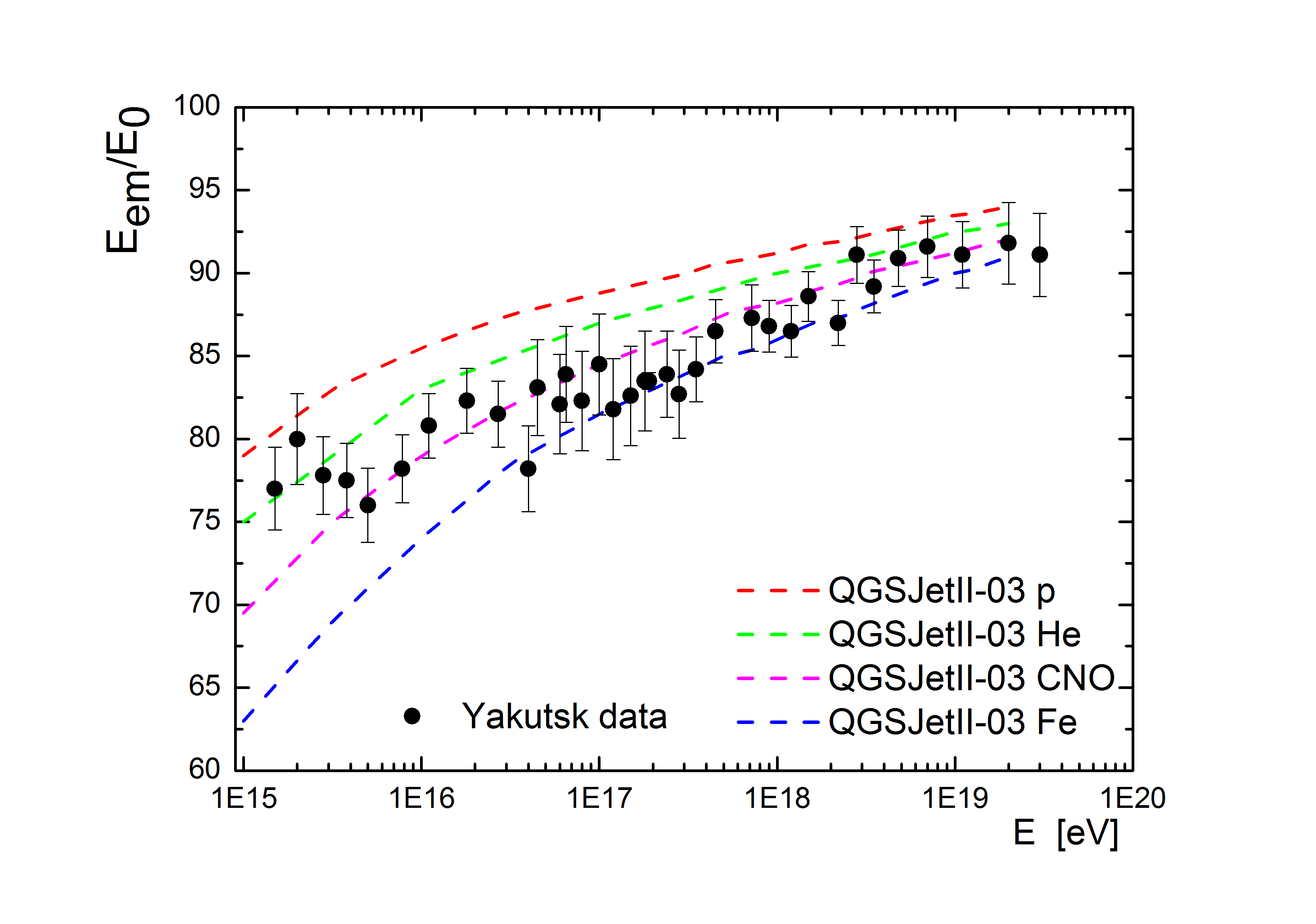}}
	\caption{The fraction of energy transferred to the electromagnetic component according to the registration of Cherenkov light of air shower at the Yakutsk array and hadronic interactions model QGSJetII-03 for proton p, helium He, CNO nuclei and iron nucleus  Fe}
	\label{fig_ykt_em_fraction}
\end{figure}

As can be seen from Fig. \ref{fig_ykt_em_fraction}, the fraction of E$_{em}$ is 85-90$ \% $, fraction of the remaining components of the shower does not exceed 10$ \% $, and their estimation is described in details in papers \citep{Knurenko2006473,Ivanov201534}. 

The experimental data of the Yakutsk array are in good agreement with the calculations by the QGSjetII-03 model in the case of a mixed composition of primary particles. This estimate may change if other models are used, including the QGSjetII-04 \citep{Ostapchenko201183}. This is stated in our work \citep{Knurenko2005206897}. In papers \citep{Abbasi2018865,Barbosa200422}, the calculations of E$_{em}$ / E$_{0}$ were performed to estimate the energy of air showers registered at the HiRes and TA arrays. Based on the calculations and the experiment in Yakutsk, it can be said that with an accuracy of 10$\% $, there is agreement on the energy estimates in both arrays.

A test check of the reconstruction method was carried out using QGSjet-01 calculations of the Cherenkov light of air showers. The result of comparing the cascade curve calculated by the model for the energy of 10$^{18}$ eV and reconstructed by the method described above showed a good agreement \cite{Knurenko2001157}. 

\subsection{Air shower reconstruction uncertainty }

Measurement errors in our experiment are systematic and random \citep{Knurenko20159904001}. The main source of systematic uncertainty is the absolute calibration error due to the uncertainty of the transition from the amplitude of the signal at the output of the PMT to the number of Cherenkov light photons. It is determined by the coupling coefficient between the amplitudes of the signals from the reference light source and the plastic scintillator unit, which serves as a calibration light source. The relative error of this coefficient is $ \delta_{Q_{cal}} $ = 21$ \% $.

Random measurement errors are due to many factors. One of them is the instrumental error of response measurement itself, equal to $ \delta_{Q_{app}} $ = 13$ \% $. For the short (R $ < $ 100 m) distances from the axis, a similar contribution is made by axis determination method, the absolute error of which for showers selected by the central part of the array is $ \Delta $R = 10 m for the approximation of Q (R) by a piecewise-power function with value of n $ \sigma_{loc} $ = n$ \cdot \Delta $ (R / R$_{thresh.}$). Along with this, there are Q (R) distortions due to the shower processing procedure. These include the use of different test functions with different boundary conditions for the density of charged particles, the classification of showers according to different measured parameters, the threshold of the Cherenkov light detector. The study of these distorting factors shows that in this case measurement errors of Q (R) increase at small and large (R$ > $ 800 m) distances from the shower axis. The influence of the detector threshold begins to affect from a distance of R$ > $ 600 m, depending on the energy of the primary particle. It is expressed in the fact that due to the random nature of the detector triggering at the threshold level, zero readings are underestimated and the value of the Cherenkov light flux begins to be systematically overestimated. For E$_{0}$ = 2$\cdot$10$^{18}$ eV, this overestimation is 5$ \% $ and 20$ \% $ at distances of 600 and 800 m, respectively, and this fact had to be taken into account when processing showers. For this reason, showers for the analysis were selected only by the central part of the array, where the symmetry condition of the location of observation stations relative to the air shower axis was met and all showers had approximately the same uncertainty for the reconstruction of the air shower characteristics.

Above mentioned uncertainties were taken into account for reconstruction of the cascade curve according to formula (\ref{yakutsk_eq4}):

\begin{equation}\label{yakutsk_eq4}
\sigma_{Q_{i}}^{2} = \left(
0.04+n^{2}
\cdot
\left(
\frac{\Delta R}
{R}\right)^{2}
\cdot 
Q^{2}_{i}
\right)
\end{equation}

Here $ \Delta $R - is error of air shower axis determination error, 0.04 - constant associated with the absolute calibration of the light receivers and n - slope value of the power approximation of Cherenkov light LDF.

Uncertainty of the X$ _{max} $ reconstruction for individual showers was 15-55 g/cm$^{2} $ \citep{Knurenko2003168d,Knurenko2001157}. Uncertainty of energy estimation at the Yakutsk array, taking into account the measurements uncertainties listed above is 23-26$ \% $ \citep{Knurenko2006473,Ivanov200911}.  

\subsection{Depth of maximum development X$_{max} $}

\begin{figure}[h]
	\center{\includegraphics[width=0.8\linewidth]{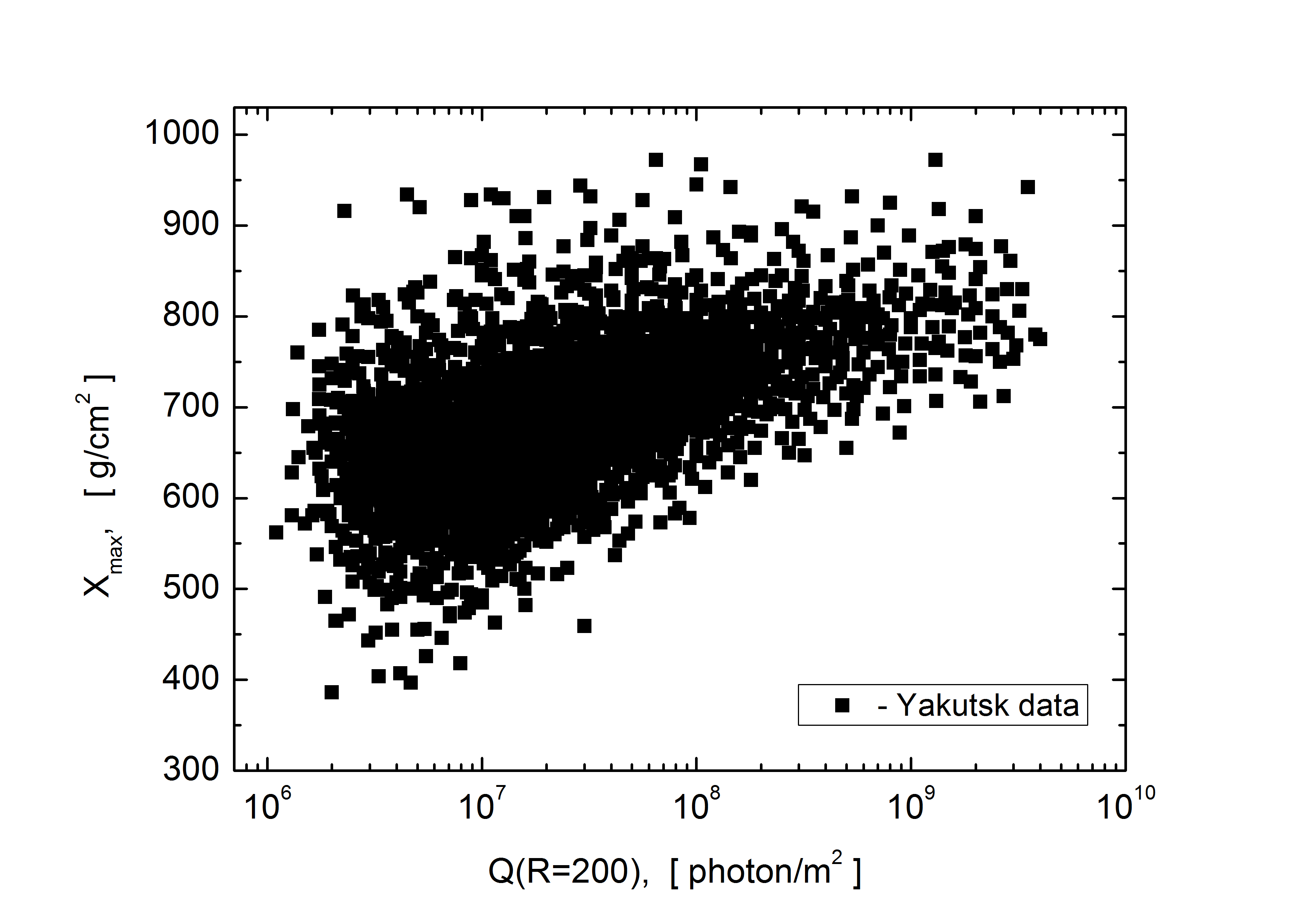}}
	\caption{Dependence of X$_{max}$ from classification parameter Q (200) - EAS Cherenkov light flux density at a distance of 200 m from the shower axis. }
	\label{fig_ykt_selected_data}
\end{figure}

The analysis presented in this paper is based on the Yakutsk array 1974-2014 dataset. The showers selected for analysis are shown in Fig. \ref{fig_ykt_selected_data}, as the dependence of X$_{max}$ on the classification parameter of air showers Q (200) - Cherenkov light flux density at the distance 200 m from the shower axis. The flux density of the Cherenkov light has little shower-to-shower fluctuations and weakly depends on zenith angle \citep{Kalmykov19676, Knurenko2003168d}, therefore Q(200) was used as the classification parameter for air shower selection. The energy of the shower is derived from Q (200) parameter by formula (\ref{yakutsk_eq5}) \citep{Knurenko2006473}: 

\begin{equation}\label{yakutsk_eq5}
E_0=(1.78\pm 0.44)\cdot 10^{17}\cdot \left(\frac{Q(200)}{10^{17}}\right)^{1.01\pm 0.04} 
\end{equation}

In the distribution of air showers given in Fig.\ref{fig_ykt_selected_data}, there are events with X$_{max}$ $\geq$ 800 g / cm$^{2}$, i.e. with the low depth of maximum development compared to iron or even proton induced air showers. These showers can be used to search for neutral particles such as $\gamma$-rays and neutrinos \citep{Knurenko2018107676}. 

\begin{figure}[h]
	\begin{minipage}[h]{0.5\linewidth}
		\center{\includegraphics[width=1.0\linewidth]{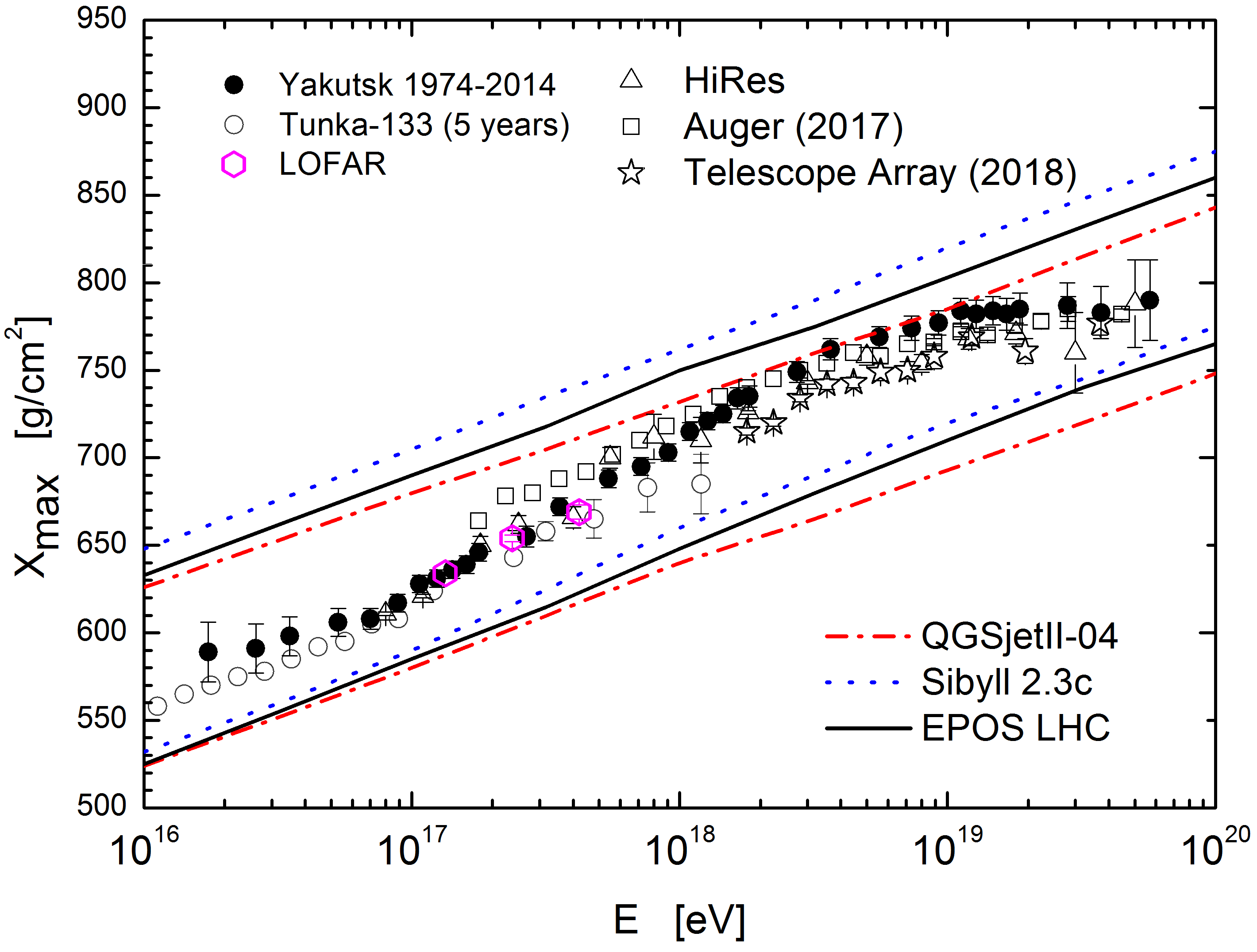} \\ a)}
	\end{minipage}
	\begin{minipage}[h]{0.5\linewidth}
		\center{\includegraphics[width=1.0\linewidth]{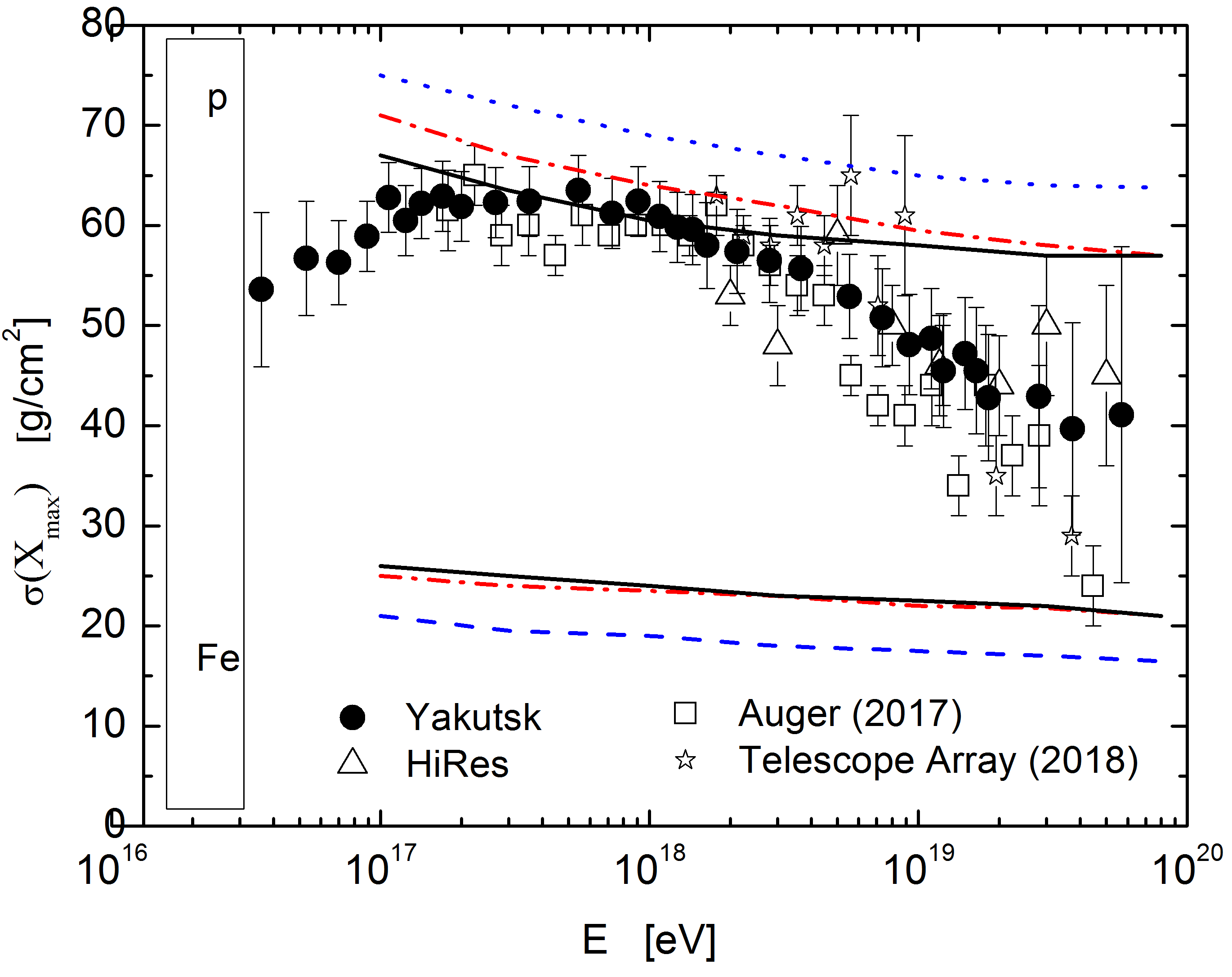} \\ b)}
	\end{minipage}
	\caption{The mean (a) and the standard deviation (b) of the measured X$_{max}$ distributions as a function of energy obtained from Cherenkov light data in 1974-2014 at the Yakutsk array. Comparison with other experiments: Auger \citep{Bellido2018506}, TA \citep{Abbasi201876}, HiRes \citep{Abbasi2010104,Ulrich201241}, Tunka \citep{Prosin2016121}, LOFAR \citep{Horandel2016531}  and simulations for proton and iron primaries by different models QGSJetII-04 \citep{Ostapchenko201183}, Sibyll 2.3c \citep{Riehn2015ANV}, EPOS LHC \citep{Pierog201592}.  }
	\label{fig_ykt_xmax}
\end{figure}

The averaged data of the Yakutsk array together with the data of the Auger \citep{Bellido2018506}, TA \citep{Abbasi201876}, HiRes \citep{Abbasi2010104,Ulrich201241}, Tunka \citep{Prosin2016121}, LOFAR \citep{Horandel2016531} arrays are shown in Fig. \ref{fig_ykt_xmax} (a) and comparison with different hadronic interaction models QGSJetII-04 \citep{Ostapchenko201183}, Sibyll 2.3c \citep{Riehn2015ANV}, EPOS LHC \citep{Pierog201592}. The data within the experimental errors are in good agreement with each other, although they were obtained at different experiments and by different methods. The data of the Yakutsk array covers almost four orders of magnitude in energy. Statistics allows with sufficient accuracy to identify the shift of X$_{max}$ with increasing energy and to estimate the elongation rate (ER) at different energy ranges. According to our data, the ER is:

\begin{enumerate}
\item E = 10$^{16}$-10$^{17}$ [eV]; E.R. =  48$\pm$6      [ g/cm$^{2}$ ] 
\item E = 10$^{17}$-10$^{18}$ [eV]; E.R. =  78$\pm$5      [ g/cm$^{2}$ ]
\item E = 10$^{18}$-10$^{19}$ [eV]; E.R. =  63$\pm$6      [ g/cm$^{2}$ ]
\item E = 10$^{19}$-10$^{20}$ [eV]; E.R. =  50$\pm$7      [ g/cm$^{2}$ ]

\end{enumerate}

The data $\langle X_{max}\rangle$, $\sigma$(X$_{max}$) and other characteristics are given in Tables \ref{yakutsk_tab1}, \ref{yakutsk_tab2}.

In Fig. \ref{fig_ykt_xmax} (b) X$_{max}$ fluctuations are presented. These fluctuations were obtained taking into account detectors effect according to the method given below. The accuracy of the X$_{max}$ reconstruction was estimated from the simulation of cascade curve reconstruction process by full Monte Carlo method, given the known values of the instrumental uncertainty Q(E$_{0}$, R, X$_{0}$) and parameters of the average cascade curve. It was assumed that the measurement uncertainty Q(E$_{0}$, R, X$_{0}$) are distributed according to the normal law \citep{Dyakonov19816, Dyakonov19833447}.  As follows from \citep{Dyakonov19816, Dyakonov19833447} $\sigma(X_{max})_{app} $ depends not only on measurements uncertainty Q(E$_{0}$, R, X$_{0}$), but also on coordinates of the shower axis relative to the array center (X$_{0} $, Y$_{0} $) and the number of stations n participating in the registration of the Cherenkov EAS light. So the average instrumental error at an energy of 10$^{18}$ eV is equal to $<\sigma(X_{max})_{app}>$ = 38 g / cm$^{2}$, and its dependence on energy can be expressed as:

\begin{equation}
\langle
\sigma (X_{max})_{app}
\rangle
=
(
38.5 \pm 5
)
-
(
10 \pm 3
)
\lg 
\frac{E_0}{10^{18}}
\end{equation}

Then physical fluctuations  $\sigma(X_{max})$ were found as the difference of measured fluctuations and instrumental fluctuations: 
\begin{equation}
\sigma^{2}(X_{max})
= 
\sigma^{2}(X_{max})_{meas}
-
\sigma^{2}(X_{max})_{app}
\end{equation}

The data was divided by energy with a step of 1.5, and the value $\sigma$(X$_{max}$) was found for each interval. The figure contains data obtained at the Yakutsk array and data of Auger, TA and HiRes.  The figure also shows calculations using modern models of hadron interactions for the primary proton and the iron nucleus.

From Fig. \ref{fig_ykt_xmax} (b) one can see there is a maximum of X$_{max}$ fluctuations dependency on energy in the range 2$\cdot$10$^{17}$-2$\cdot$10$^{18}$ eV. In this range, the value of $\sigma$(X$_{max}$) is 57-63 g/cm$^{2}$, which according to model calculations corresponds to the mixed composition of cosmic rays with a high content of p + He and heavier nuclei. However, the composition consisting of a mix of 50{\%} proton p and 50${\%}$ iron nuclei Fe cannot be excluded.  A decrease in fluctuations above the energy of 5$\cdot$10$^{18}$ eV indicates a heavier composition.

\subsection{Mass composition}

Knowing the depth of the shower maximum X$_{max}$ (Fig. $\ref{fig_ykt_xmax}$, a) and average X$_{max}$ for the proton and the iron nucleus from QGSJetII-04 simulations, the value of $\langle lnA \rangle$ can be determined by interpolation using formula (\ref{yakutsk_eq6}) \citep{Berezhko201231,Horandel200641}:

\begin{equation}\label{yakutsk_eq6}
\langle lnA \rangle = 
\frac{X^{exp}_{max} - X^{p}_{max}}
{X^{Fe}_{max} - X^{p}_{max}}
\cdot lnA_{Fe}
\end{equation}

where X$^{exp}_{max}$ is the depth of maximum determined from the experiment, $lnA_{Fe}$ is the natural logarithm of the iron atomic mass.

\begin{figure}[h]
	\center{\includegraphics[width=0.8\linewidth]{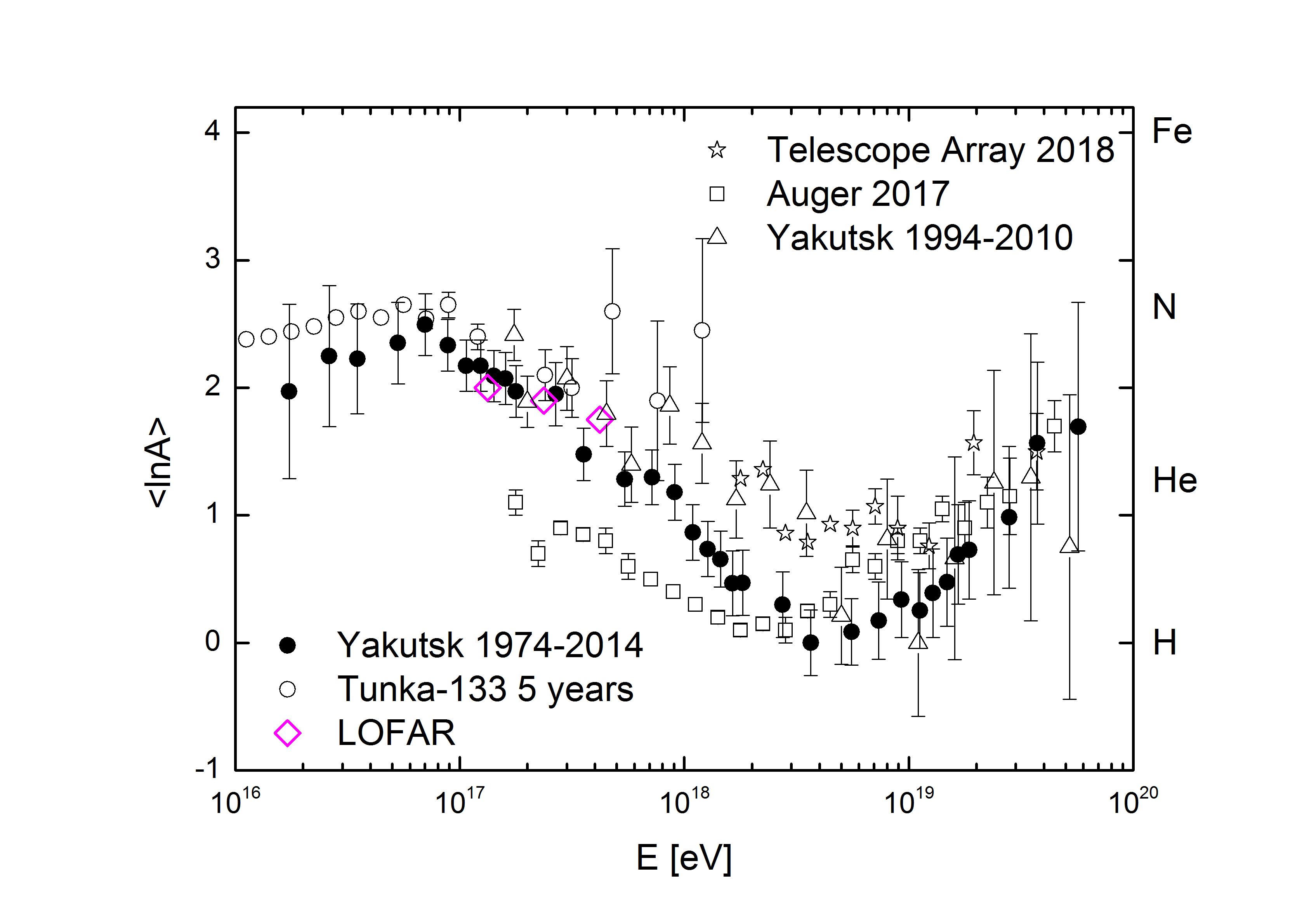}}
	\caption{Average atomic mass $\langle lnA \rangle$ derived from the average depth of the shower maximum with QGSJetII-04 \citep{Ostapchenko201183}. Comparison with Yakutsk (1994-2010)\citep{Knurenko20117251}, Tunka \citep{Prosin2016121}, LOFAR \citep{Horandel2016531}, TA \citep{Abbasi201999} and Auger \citep{Bellido2018506} data.  }
	\label{fig_ykt_CR_MC}
\end{figure}

The mean logarithmic mass values are presented for different energies in Fig. \ref{fig_ykt_CR_MC}. The results are compared with data of Tunka \citep{Prosin2016121}, LOFAR \cite{Horandel2016531}, TA \citep{Abbasi201999} and Auger \citep{Bellido2018506}, derived using the hadronic interaction model QGSJetII-04. There is also the Yakutsk array data for 1994-2010 \citep{Knurenko20117251}]. This result was obtained for limited data selection with QGSJETII model, which explains discrepancy with the data presented in current paper As can be seen from the figure, there are irregularities of the value of $\langle lnA\rangle$, due to MC change. For example, in the energy region of 10$^{16}$-10$^{17}$ eV, the composition is heavier than in the energy region of 10$^{17}$-10$^{18}$ eV, where composition mostly consists of protons p and helium He nuclei. Starting from $\sim$10$^{19}$ eV, the composition becomes heavier, i.e. the fraction of such nuclei as CNO and Fe iron nuclei increases.

\begin{table}
	\centering
	\caption{$\left\langle X_{max} \right\rangle$ , $\sigma$( X$_{max}$ ) and $\left\langle lnA \right\rangle $ observed in 1974-2014 at the Yakutsk array.10$^{16}$ - 10$^{18}$ eV }
	\label{yakutsk_tab1}       
	\begin{tabular}{|c|c|c|c|c|c|}
		\hline
		E$_{0}$, eV & N, events &X$_{max}$, g/cm$^{2}$ & $\sigma$(X$_{max}$), g/cm$^{2}$  & $\langle lnA \rangle$ & $\langle lnA \rangle$ stat. error \\ \hline
		1.74E+16   &   -       &589$\pm$17            &                -                   &         1.97          &               0.68                \\ \hline
		2.62E+16   &   -       &591$\pm$14            &                -                   &         2.25          &               0.55                \\ \hline
		3.50E+16   &   24      &598$\pm$11            &              53.6$\pm$7.7          &         2.23          &               0.43                \\ \hline
		5.31E+16   &   50      &606$\pm$8       	   &             56.7$\pm$5.7          &         2.35          &               0.32                \\ \hline
		7.01E+16   &   88      &608$\pm$6       	   &              56.3$\pm$4.2          &         2.50          &               0.24                \\ \hline
		8.85E+16   &   139     &617$\pm$5       	   &              58.9$\pm$3.5          &         2.34          &               0.20                \\ \hline
		1.07E+17   &   158     &628$\pm$5       	   &              62.8$\pm$3.5          &         2.17          &               0.20                \\ \hline
		1.24E+17   &   146     &631$\pm$5       	   &              60.5$\pm$3.5          &         2.17          &               0.20                \\ \hline
		1.42E+17   &   155     &636$\pm$5       	   &              62.2$\pm$3.5          &         2.09          &               0.20                \\ \hline
		1.60E+17   &   158     &639$\pm$5       	   &              62.9$\pm$3.5          &         2.07          &               0.20                \\ \hline
		1.78E+17   &   153     &646$\pm$5       	   &              61.9$\pm$3.5          &         1.97          &               0.20                \\ \hline
		2.68E+17   &   155     &655$\pm$5       	   &              62.3$\pm$3.5          &         1.95          &               0.25                \\ \hline
		3.57E+17   &   156     &672$\pm$5       	   &              62.4$\pm$3.5          &         1.48          &               0.21                \\ \hline
		5.43E+17   &   161     &688$\pm$5       	   &              63.5$\pm$3.5          &         1.28          &               0.21                \\ \hline
		7.18E+17   &   150     &695$\pm$5       	   &              61.2$\pm$3.5          &         1.30          &               0.22                \\ \hline
		9.06E+17   &   155     &703$\pm$5       	   &              62.4$\pm$3.5          &         1.18          &               0.22                \\ \hline
	\end{tabular}
\end{table}

\begin{table}
	\centering
	\caption{$\left\langle X_{max} \right\rangle$ , $\sigma$( X$_{max}$ ) and $\left\langle lnA \right\rangle $ observed in 1974-2014 at the Yakutsk array. 10$^{18}$ $\cdot$5.7$\cdot$10$^{19}$ eV}
	\label{yakutsk_tab2}       
	\begin{tabular}{|c|c|c|c|c|c|}
		\hline
		E$_{0}$, eV & N, events &X$_{max}$, g/cm$^{2}$ & $\sigma$(X$_{max}$), g/cm$^{2}$ & $\langle lnA \rangle$ & $\langle lnA \rangle$ stat. error \\ \hline
		1.09E+18   & 148       &715$\pm$5       &              60.9$\pm$3.5         &         0.87          &               0.22                \\ \hline
		1.27E+18   & 143       &721$\pm$5       &              59.8$\pm$3.5         &         0.73          &               0.22                \\ \hline
		1.45E+18   & 142       &725$\pm$5       &              59.6$\pm$3.5         &         0.66          &               0.22                \\ \hline
		1.64E+18   & 93        &734$\pm$6       &              58.0$\pm$4.3         &         0.47          &               0.25                \\ \hline
		1.82E+18   & 92        &735$\pm$6       &              57.4$\pm$4.2         &         0.47          &               0.26                \\ \hline
		2.74E+18   & 89        &749$\pm$6       &              56.5$\pm$4.2         &         0.30          &               0.26                \\ \hline
		3.66E+18   & 86        &762$\pm$6       &              55.7$\pm$4.2         &           0           &               0.26                \\ \hline
		5.56E+18   & 78        &769$\pm$6       &              52.9$\pm$4.2         &         0.09          &               0.26                \\ \hline
		7.35E+18   & 53        &774$\pm$7       &              50.8$\pm$4.9         &         0.17          &               0.30                \\ \hline
		9.27E+18   & 47        &777$\pm$7       &              48.1$\pm$5.0         &         0.34          &               0.30                \\ \hline
		1.12E+19   & 48        &784$\pm$7       &              48.7$\pm$5.0         &         0.25          &               0.30                \\ \hline
		1.28E+19   & 32        &782$\pm$8       &              45.5$\pm$5.7         &         0.39          &               0.35                \\ \hline
		1.48E+19   & 35        &784$\pm$8       &              47.2$\pm$5.6         &         0.48          &               0.35                \\ \hline
		1.66E+19   & 26        &782$\pm$9       &              45.5$\pm$6.3         &         0.69          &               0.39                \\ \hline
		1.86E+19   & 23        &785$\pm$9       &              42.8$\pm$6.3         &         0.73          &               0.39                \\ \hline
		2.80E+19   & 11        &787$\pm$13      &              42.9$\pm$9.1         &         0.99          &               0.56                \\ \hline
		3.74E+19   & 7         &783$\pm$15      &              39.7$\pm$10.6        &         1.57          &               0,64                \\ \hline
		5.69E+19   & 3         &790$\pm$23      &              41.1$\pm$16.8        &         1.69          &               0.97                \\ \hline
	\end{tabular}
\end{table}

\section{Conclusion}

The Yakutsk array is operating continuously for more than 45 years, detecting electrons, muons, Cherenkov light, and radio emission. At the same time, more than 5$\cdot$10$^{6}$ EAS events were recorded in the energy region above 10$^{15}$ eV. Using a large dataset of air showers with energies of 10$^{16}$-5.7$\cdot$10$^{19}$ eV, registered for the period 1974-2014, the longitudinal development of showers was reconstructed and the $X_{max}$ dependence was found. The results showed that elongation rate per decade of energy has nonlinear advancement and takes the values 48$\pm$6, 78$\pm$5, 63$\pm$6, 50$\pm$7 g / cm$^{2}$ for energy ranges, 10$^{16}$-10$^{17}$, 10$^{17}$-10$^{18}$, 10$^{18}$-10$^{19}$, 10$^{19}$-10$^{20}$ eV. As can be seen (Fig. \ref{fig_ykt_xmax}), there are kinks at the energies $\sim$10$^{17}$ eV and $\sim$5$\cdot$10$^{18}$ eV, i.e. on the "second knee" area and the beginning of irregularities in the spectrum like "dip-bump". As shown by the data of the Yakutsk array (Fig. $\ref{fig_ykt_CR_MC}$), this is due to a change in the cosmic ray mass composition. At low energies up to 10$^{17}$ eV, there are noticeably more nuclei with an atomic mass of 4-56, at energies of 10$^{17}$-10$^{18}$ eV the proportion of protons increases and reaches a maximum, accounting for 60-80$\%$. In the region of energy greater than 10$^{19}$ eV, CR predominantly consist of helium nuclei He, CNO, and heavier elements.
\section*{Acknowledgement} 
The reported study was funded by RFBR according to the research project 16-29-13019.

\section*{References}

\bibliography{mybibfile}

\end{document}